\documentclass[aps,prl,twocolumn,longbibliography]{revtex4-2}%
\usepackage{graphicx}
\usepackage{float}
\usepackage{dcolumn}
\usepackage{bm,color}
\usepackage{amsmath,amssymb,dsfont,amstext,amsfonts}
\usepackage{extarrows}
\usepackage[colorlinks=true,linkcolor=blue,urlcolor=blue,citecolor=blue]{hyperref}
\usepackage{xcolor}
\usepackage{wasysym} 
\usepackage{mathtools}
\usepackage{bbold} 
\usepackage{microtype}
\usepackage{braket}
\usepackage{comment}
\usepackage[export]{adjustbox}% for valign in includegraphics
\usepackage[caption=false,subrefformat=parens,labelformat=empty]{subfig}
\usepackage{dsfont}

\usepackage[normalem]{ulem} %makes underlining possible: \uline \uwave \sout
\usepackage{color}%makes \textcolor{grey}{text} available

\let\oldciteauthor=\citeauthor
\def\citeauthor#1{\hypersetup{citecolor=blue}\oldciteauthor{#1}}

\let\oldciten=\onlinecite
\def\onlinecite#1{\hypersetup{citecolor=blue}\oldciten{#1}}

\let\oldcite=\cite
\def\cite#1{\hypersetup{citecolor=blue}\oldcite{#1}}

\begin{document}
\title{Witnessing Short- and Long-Range Nonstabilizerness via the Information Lattice}

\author{Yuliya Bilinskaya}\thanks{These, alphabetically listed, authors contributed equally to this work.}
\affiliation{Department of Physics, KTH Royal Institute of Technology, 106 91, Stockholm, Sweden}
\author{Miguel F.\ Martínez}\thanks{These, alphabetically listed, authors contributed equally to this work.}
\affiliation{Department of Physics, KTH Royal Institute of Technology, 106 91, Stockholm, Sweden}
\author{Soumi Ghosh}
\affiliation{Department of Physics, KTH Royal Institute of Technology, 106 91, Stockholm, Sweden}
\author{Thomas~Klein~Kvorning}
\affiliation{Department of Physics, KTH Royal Institute of Technology, 106 91, Stockholm, Sweden}
\author{Claudia Artiaco}
\affiliation{Department of Physics, KTH Royal Institute of Technology, 106 91, Stockholm, Sweden}
\author{Jens H. Bardarson}
\affiliation{Department of Physics, KTH Royal Institute of Technology, 106 91, Stockholm, Sweden}

\begin{abstract}
We study nonstabilizerness on the information lattice and demonstrate that noninteger local information directly indicates nonstabilizerness. 
For states with a clear separation of short- and large-scale information, a noninteger total information at large scales $\Gamma$ serves as a witness of long-range nonstabilizerness. 
We propose a folding procedure to separate the global and edge-to-edge contributions to $\Gamma$.
As an example we show that the ferromagnetic ground state of the spin-1/2 three-state Potts model has long-range nonstabilizerness originating from global correlations, while the paramagnetic ground state has at most short-range nonstabilizerness.
\end{abstract}

\maketitle
\textit{Introduction}---Quantum information theory offers a unifying lens for understanding quantum states and dynamics~\cite{srednicki1993entropy, hastings2007area, wolf2008area, eisert2010colloquium, deutsch1991quantum,  srednicki1994chaos, rigol2008thermalization, laflorencie2016quantum, zeng2019quantum}. 
In the context of condensed matter, states are often classified in terms of the scaling of the subsystem von Neumann entropy, which is a global measure of the total correlations between a subsystem and its complement.
In area-law states~\cite{srednicki1993entropy, hastings2007area, wolf2008area, eisert2010colloquium}, such as ground states of local gapped Hamiltonians and many-body localized states, correlations are mainly local, meaning that the von Neumann entropy of a subsystem scales with its area.
This enables efficient representation of such states via matrix product states~\cite{eisert2010colloquium, schollwock2011thedensitymatrix, cirac2021matrix}.
In contrast, volume-law states~\cite{deutsch1991quantum, srednicki1994chaos, rigol2008thermalization}, such as midspectrum eigenstates of ergodic Hamiltonians, have correlations involving an extensive number of degrees of freedom. 
Furthermore, constant subleading corrections to the area-law scaling serve to characterize states with topological order~\cite{hamma2005ground, kitaev2006topological, levin2006detecting, chen2010local, masmendoza2025graphical}. 

Quantum states can also be characterized by other quantum information properties such as nonstabilizerness.
Stabilizer states are closed under Clifford operations and admit an efficient classical representation~\cite{knill1996nonbinary, gottesman1998heisenberg}.
Generic condensed matter states require non-Clifford unitaries to be reached from a stabilizer state.
Such unitaries are constructed from a universal set realized, for example, by the combination of Clifford and $T$-gates.
The deviation of a state from a stabilizer state is known as nonstabilizerness, or magic, and has been quantified by different measures~\cite{emerson2013theresource, howard2017application, bravyi2019simulation, leone2022stabilizer, jiang2023lower, tirrito2024quantifying, dowling2025magic, dowling2025bridging, paviglianiti2025estimating}.
While the concept of nonstabilizerness arose in the context of quantum error correction and fault-tolerant quantum computation~\cite{gottesman1997stabilizer, gottesman1998theory, bravyi2005universal, fowler2012surface}, it has increasing relevance in condensed matter and quantum many-body physics~\cite{sarkar2020characterization, oliviero2022magicstate, rattasaco2023stabilizer, passarelli2024nonstabilizerness, tarabunga2024critical, bejan2024dynamical, fux2025disentangling, turkeshi2025pauli, passarelli2025chaos, tirrito2025anticoncentration, tirrito2025magic, russomanno2025nonstabilizerness, odavic2025stabilizer, viscardi2025interplay, collura2025nonstabilizerness, sarkis2025aremolecules, turkeshi2025magic, falcao2025magic, hernandez2025non}.

Nonstabilizerness has been interpreted as a measure of quantum-state complexity~\cite{santra2025complexity, haug2025probing, collura2025nonstabilizerness, sierant2025fermionic, huang2025nonstabilizerness, fux2025disentangling} as it quantifies the distance from the easily representable stabilizer states.
Other measures based on the distance from other classes of states that admit efficient representations, such as area-law or fermionic Gaussian states~\cite{valiant2001quantum, terhal2002classical, sierant2025fermionic}, provide similar yet inequivalent quantifications of complexity.
The existence of such different classes raises the requirement of a broader classification.
For instance, a comprehensive characterization of quantum states can be obtained by looking at the scale of both correlations and nonstabilizerness. 

From the point of view of correlations, the information lattice~\cite{klein2022time, artiaco2025universal} provides a universal characterization of quantum states in terms of local information, which quantifies the total amount of correlations per scale $\ell$ and spatial location $n$.
The information lattice defines the intrinsic correlation length scales of a state, which do not depend on any specific observable~\cite{artiaco2025universal}. 
Within the information lattice, local information behaves as a hydrodynamic quantity: under the application of a local quantum gate information can only flow between nearby lattice points and there exist well-defined information currents~\cite{klein2022time}.
This allows us to comprehensively characterize quantum quench dynamics~\cite{bauer2025local,artiaco2025out,barata2025hadronic,nava2026information} and to perform efficient time evolution of large-scale quantum many-body systems~\cite{klein2022time,artiaco2024efficient,harkins2025nanoscale}. 
As opposed to information,  there is no well-defined decomposition indicating where and on what scale nonstabilizerness resides, but there are attempts to evaluate nonlocality of nonstabilizerness~\cite{tarabunga2023manybody, qian2025quantum, cao2025nontrivial, cao2025gravitational, tarabunga2025efficient, timsina2025robustness}, as well as to divide nonstabilizerness into short- and long-range~\cite{white2021conformal, ellison2021symmetry,korbany2025longrange}. 
The latter approach relies on an equivalence relation up to shallow local unitary quantum circuits as is done in topological classification~\cite{chen2010local}.
This distinction between short- and long-range nonstabilizerness is not captured by just computing a nonstabilizer monotone, which only gives a measure of the total amount of nonstabilizerness in a quantum state.
\begin{figure}[tb]
    \includegraphics[width=\columnwidth]{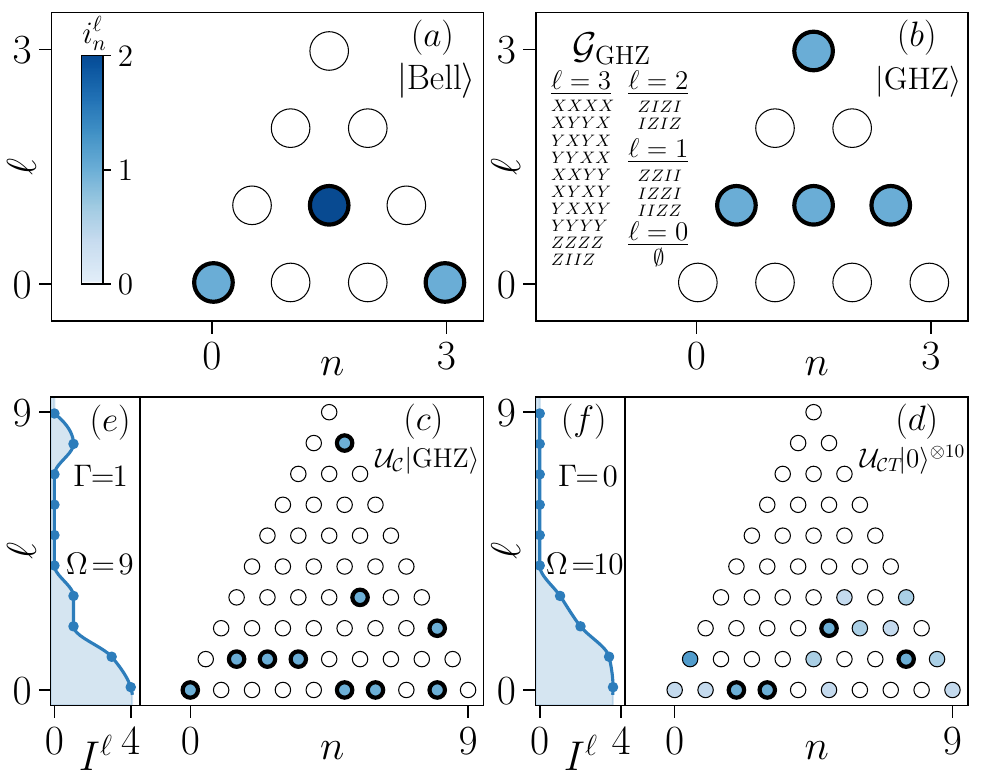}
    \caption{
    Information lattice of (a) $\ket{\rm Bell}=\frac{1}{\sqrt{2}}(\ket{0101}+\ket{0011})$, and (b) $\ket{\rm GHZ}=\frac{1}{\sqrt{2}}(\ket{0000}+\ket{1111})$ with its stabilizer group $\mathcal{G}_{\rm GHZ}$ (excluding the identity operator).
    (c) Information lattice of a stabilizer state constructed by applying a 2-layer random Clifford circuit $\mathcal{U}_{\mathcal{C}}$ to the ten-qubit GHZ state. 
    (d) Information lattice of a short-range nonstabilizer state constructed by applying a random $T$-doped Clifford circuit $\mathcal{U}_{\mathcal{C}T}$ on the product state $\ket{0}^{\bigotimes 10}$. 
    The circuit $\mathcal{U}_{\mathcal{C}T}$ is composed of three blocks, each one containing $10$ layers of Clifford gates and $5$ $T$-gates acting on randomly selected qubits and between two randomly selected Clifford layers. 
    (e, f) Information per scale $I^\ell$ of (c) and (d), respectively. 
    In (a)-(d), the color of the sites on the information lattice indicates the amount of local information $i_{n}^\ell$ in $\mathcal{C}_n^\ell$, while sites with integer values of $i_{n}^\ell$ are circled by bold black lines.
    }
    \label{fig:Info_lattice}
\end{figure}

In this work we show that any noninteger value of local information on the information lattice directly indicates nonstabilizerness in a quantum state.
Furthermore, in states where there is a clear separation between short- and large-scale local information (we refer to such states as localized), we use the information lattice to define a witness of long-range nonstabilizerness via the total information at large scales $\Gamma$~\cite{artiaco2025universal}.
Through a folding procedure, we define a coarse-grained information lattice that allows us to distinguish whether the long-range nonstabilizerness originates from global or edge-to-edge correlations.
We exemplify our approach using the spin-1/2 three-state Potts model and characterize its ground states in the paramagnetic and ferromagnetic phases in terms of their nonstabilizerness.
While the paramagnetic ground state has at most short-range nonstabilizerness, the ferromagnetic ground state exhibits long-range nonstabilizerness originating from global correlations.

\textit{Stabilizer states and the information lattice}---We consider an $L$-qubit chain with Hilbert space $\mathcal{H}=(\mathbb{C}^2) ^{\otimes L}$. 
We denote by $\mathcal{C}_n^\ell$ the subsystem of $\ell + 1$ contiguous sites centered at position $n$, and refer to $\ell$ as the scale of $\mathcal{C}_n^\ell$. 
The information lattice of a state $\rho$ acting on $\mathcal{H}$ is a two-dimensional structure where subsystems are ordered hierarchically in terms of location $n$ and scale $\ell$~\cite{klein2022time, artiaco2025universal}.
Each site in this structure is uniquely specified by the pair $(n, \ell)$, where $n$ is the integer for even $\ell$ and half-integer for odd $\ell$,  and is associated with the local information $i_n^\ell$ in subsystem $\mathcal{C}_n^\ell$.
The local information is the maximal amount of information one can obtain by measuring any set of operators that cannot be accessed from any smaller-scale subsystem of $\mathcal{C}_{n}^\ell$~\cite{klein2022time, artiaco2025universal}.
For $\ell > 1$, the local information in subsystem $\mathcal{C}_n^\ell$ is the mutual information of its two largest subsystems, namely $\mathcal{C}_{n-1/2}^{\ell-1}$ and $\mathcal{C}_{n+1/2}^{\ell-1}$, conditioned on their overlap, which is subsystem $\mathcal{C}_{n}^{\ell-2}$ (see End Matter for details).
Formally, the local information $i^\ell_n$~\cite{klein2022time} can be written as
\begin{equation}
\begin{split}
    i_n^{\ell} = & I(\rho_n^{\ell}) - I(\rho_{n-1/2}^{\ell-1}) - I(\rho_{n+1/2}^{\ell-1}) + I(\rho_n^{\ell-2}),
\end{split}
\label{eq:local_info_subsyst}
\end{equation}
where $I(\rho_n^{\ell})= (\ell+1) - S(\rho_n^{\ell})$ is the total information with $S(\rho_n^{\ell})=-\mathrm{Tr}\left[\rho_n^{\ell} \log_2(\rho_n^{\ell})\right]$ the von Neumann entropy of $\rho_n^\ell$, the reduced density matrix of subsystem $\mathcal{C}_{n}^{\ell}$.
The local information is strictly positive and bounded by two from above, $i_n^\ell \in [0,2]$.
The information lattice thereby provides a decomposition of the total information $I(\rho)=\sum_{(n, \ell)}i_n^\ell$ into local information $i_n^\ell$.

Stabilizer states are characterized by featuring only integer values of local information, owing to their particular entanglement structure~\cite{fattal2004entanglement,sharma2025multipartite}.
A stabilizer state is the simultaneous $+1$ eigenvector of $2^L$ Pauli strings, which together form an Abelian group called the stabilizer group.
Every stabilizer group is generated by $L$ independent Pauli strings.
For a stabilizer state, the operators that yield a maximal amount of information in every subsystem are given by a particular set of stabilizer generators that we call the maximally local generating set~\cite{sharma2025multipartite}.
The elements of this set are found at each scale, starting from $\ell = 0$, by selecting the maximum number of independent Pauli strings on that scale belonging to the stabilizer group.
This construction continues until there are $L$ independent Pauli strings in the set.
The measurement of each of these operators answers a binary question corresponding to one bit of information---namely, whether or not the state is stabilized by the operator---which results in an integer local information for every subsystem.

In Fig.~\ref{fig:Info_lattice}(a) we show the information lattice of the state $\ket{\text{Bell}} = \frac{1}{\sqrt{2}}(\ket{0101} + \ket{0011})$ in which the two center qubits form a Bell pair while each of the two edge qubits is in a product state with the rest.
The maximally local generating set for the edge qubits is given by $\{ZIII, -IIIZ\}$.
In this case, the measurement of each of these operators answers whether the qubit where the operator acts nontrivially is in state $\ket{0}$ or $\ket{1}$.
The information gained by each of these binary questions is associated with the corresponding site on the information lattice, yielding a local information $i^0_0 = i^0_3 = 1$ fully contained at scale $\ell=0$.
The two bits of information associated with the maximally entangled pair at the center are only accessible by measuring any given pair of two-qubit operators in the set $\{IXXI, IYYI, -IZZI\}$.
This is reflected in the information lattice of the state, where $i^0_1 = i^0_2 = 0$ and $i^1_{1.5} = 2$.

Another example is the $4$-qubit GHZ state, $\ket{\mathrm{GHZ}}
=\frac{1}{\sqrt{2}}(\ket{0000} + \ket{1111})$, whose information lattice and stabilizer group are given in Fig.~\ref{fig:Info_lattice}(b).
The local information of the subsystems at scale $\ell=1$ is $i_{n}^1 = 1$ and is associated with the measurement of the stabilizer operators at that scale: $ZZII$,  $IZZI$, and $IIZZ$.
These operators are independent of each other and belong to the maximally local generating set.
At scale $\ell=2$ there are no stabilizer operators independent from the ones at $\ell=1$, yielding $i_1^2=i_3^2=0$.
The remaining bit of information $i_{1.5}^3 = 1$ is located at the largest scale and can only be accessed by measuring an operator that has nontrivial support on the scale of the entire system. 
There are ten possible stabilizer operators at the largest scale, and only eight are independent from the stabilizer operators at $\ell=1$.
Choosing one of them, for example $XXXX$, completes a maximally local generating set for the GHZ state.
In general, as demonstrated in this and the previous example, the maximally local generating set may not be unique.
However, the locality of its operators, dictated by the position of the nonzero local information on the information lattice, is.

 For a stabilizer state, the von Neumann information of subsystem $\mathcal{C}_n^\ell$ is given by $I(\rho_n^{\ell}) =  \vert \mathcal{G}_n^\ell\vert \leq \ell + 1$, where $\lvert \cdot \rvert$ denotes the rank of a group~\cite{fattal2004entanglement}, defined as the number of elements in its generating set (see End Matter for details).
 $\mathcal{G}_n^{\ell}$ is the stabilizer subgroup of subsystem $\mathcal{C}_{n}^{\ell}$, which contains the elements of the stabilizer group with trivial support in the complement of $\mathcal{C}_{n}^{\ell}$.
Consequently, the local information in Eq.~(\ref{eq:local_info_subsyst}) becomes
\begin{equation}
\begin{split}
    i_n^{\ell} 
    =&\vert\mathcal{G}_n^\ell\vert-\vert\mathcal{G}_{n-1/2}^{\ell-1}\vert-\vert\mathcal{G}_{n+1/2}^{\ell-1}\vert+\vert\mathcal{G}_{n}^{\ell-2}\vert,
\end{split}
\label{eq:local_info_stab}
\end{equation}
which is integer-valued and corresponds to the number of independent generators acting nontrivially only at scale $\ell$.
This is consistent with the fact that local information is bounded by two from above: increasing a subsystem by one site adds at most two additional independent generators to the maximally local generating set.

\textit{Short- and long-range nonstabilizerness via the information lattice}---A quantum state has nonstabilizerness if it is not a stabilizer state.
Since a stabilizer state necessarily has $i_n^{\ell} \in \mathbb{N}$, a state with any noninteger value of $i_n^{\ell}$ has nonstabilizerness.
This is exemplified in Figs.~\ref{fig:Info_lattice}(c) and ~\ref{fig:Info_lattice}(d), where we show the information lattice of a stabilizer state constructed by applying a random Clifford circuit to the ten-qubit GHZ state, and that of a nonstabilizer state constructed by applying a $T$-doped Clifford circuit to the product state $\ket{0}^{\otimes10}$.
While having noninteger local information is a sufficient condition for a state to have nonstabilizerness, it is not a necessary condition (see End Matter for an example).

We classify nonstabilizerness in terms of its locality by distinguishing short- and long-range nonstabilizer states~\cite{white2021conformal, ellison2021symmetry,korbany2025longrange}.
A state has short-range nonstabilizerness if it can be transformed into a stabilizer state by a shallow local unitary circuit, while a long-range nonstabilizer state requires a local unitary circuit that scales with the system size.
To capture these two classes of nonstabilizerness we use the information lattice, which has been shown to classify quantum states in terms of the information per scale,
\begin{equation}
I^{\ell} = \sum_n i_n^{\ell}.
\end{equation}

We are interested in localized states, which exhibit an extensive amount of information per scale at $\ell \ll L$, captured by 
\begin{equation}
\Omega=\sum_{\ell=0}^{\lfloor L/2 -1\rfloor} I^{\ell}.
\end{equation}
Examples of such states are the ground states of local Hamiltonians and eigenstates of many-body localized systems.
Some localized states may also have an intensive $\mathcal{O}(1)$ contribution to local information at scales $\ell \sim L$.
This contribution may originate from edge-to-edge correlations, as for example in symmetry-protected topological states, or global correlations, as in cat states~\cite{artiaco2025universal}.
In localized states, the information per scale $I^{\ell}$ decays exponentially away from the information peaks both at short and large scales, creating an information gap, see Figs.~\ref{fig:Info_lattice}(e) and ~\ref{fig:Info_lattice}(f).
While the information gap is extensive, the $\mathcal{O}(1)$ contribution at large scales stays constant and is captured by the total information at large scales
\begin{equation}
    \Gamma = \sum_{\ell= \lfloor L/2 \rfloor}^{L-1} I^{\ell},
    \label{eq:gamma}
\end{equation}
which serves as a universal characteristic of topological phases~\cite{artiaco2025universal}.
In Figs.~\ref{fig:Info_lattice}(e) and ~\ref{fig:Info_lattice}(f) we show the total information per scale of states in Figs.~\ref{fig:Info_lattice}(c) and ~\ref{fig:Info_lattice}(d) and calculate the corresponding values of $\Gamma$ and $\Omega$. 

Under a local unitary circuit the information flow on the information lattice is strictly local.
As a result the only way to close the information gap and redistribute information between $\Omega$ and $\Gamma$ is by applying a circuit that scales with the system size.
For an information gap extending over $g$ scales, the minimum depth of such a circuit is roughly $g/4$, thus $\sim \mathcal{O}(L)$~\cite{klein2022time, artiaco2024efficient}.
Consequently, a localized state with noninteger $\Gamma$ cannot be transformed into a stabilizer state, which always has integer $\Gamma$, by any shallow local unitary circuit.
Thus, $\Gamma \notin \mathbb{N}$ is a witness of long-range nonstabilizerness for localized states.
This implies that a localized state with $\Gamma = 0$ can have at most short-range nonstabilizerness, as for example is the case for the state in Fig.~\ref{fig:Info_lattice}(d).

In localized states in the thermodynamic limit, the total information at large scales $\Gamma$ separates into contributions coming from edge-to-edge and global correlations, such that $\Gamma =: \gamma_{\rm edge} + \gamma_{\rm global}$.
To estimate $\gamma_{\rm global}$ in a given state, we employ a folding procedure in which sites $n$ and $L-n-1$ are combined, reducing the chain length by half, while doubling the local Hilbert space dimension at each site. 
Because of this coarse graining, the edge-to-edge correlations of the initial chain become local on the information lattice of the folded chain.
Therefore, only the global correlations survive at large scales and the total information at large scales of the folded information lattice $\Gamma_{\rm folded}$, calculated according to Eq.~\ref{eq:gamma} with $L \rightarrow L'=L/2$, provides an estimate of $\gamma_{\rm global}$
(see End Matter).
In the next section, we show that the symmetric ground state of the spin-1/2 three-state Potts model realizes long-range nonstabilizerness. 
\begin{figure}[tb]
    \centering
    \includegraphics[width=\columnwidth]{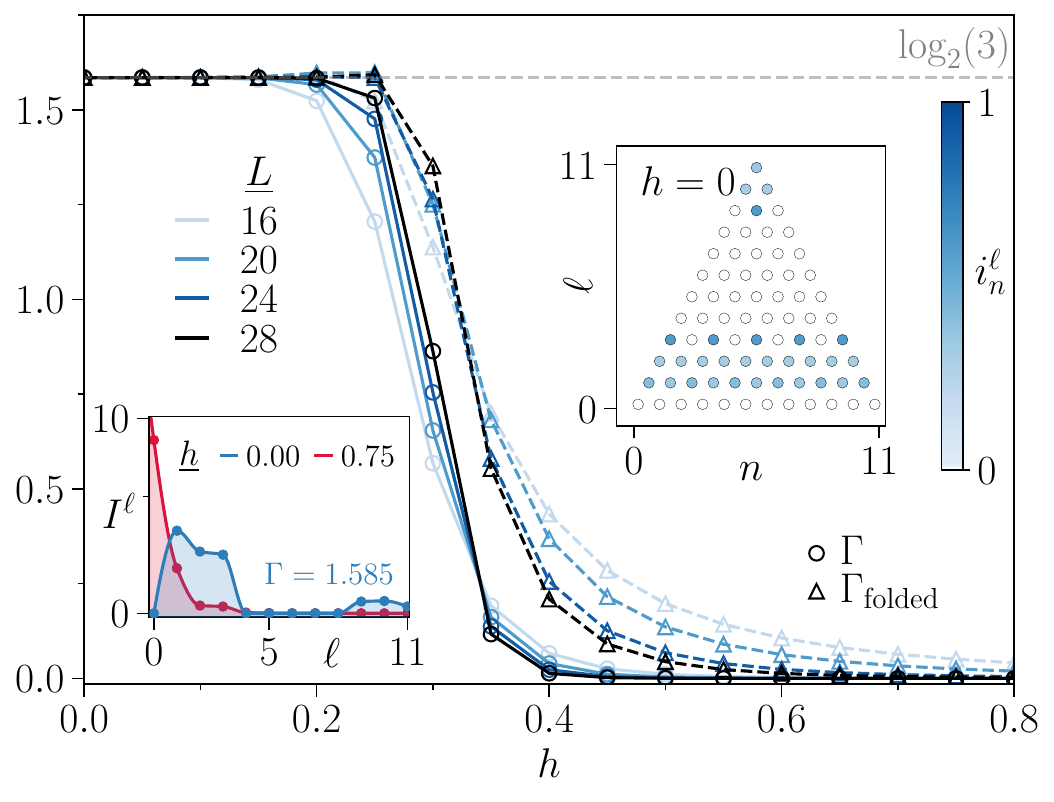}
    \caption{
    The total information at large scales in the unfolded and folded chain of the symmetric ground state of the spin-$1/2$ three-state Potts model, as a function of the magnetic field $h$ and system size $L$, with $J=1$.
    The inset on the right shows the information lattice of the symmetric ground state at $h=0$ and $L=12$. 
    The left inset shows the information per scale for the same state (blue line) and for the paramagnetic ground state at $h=0.75$ and $L=12$ (red line). 
    }
    \label{fig:Potts}
\end{figure}

\textit{Spin-1/2 three-state Potts model}---The three-state Potts model~\cite{wu1982thepotts} is described by the Hamiltonian 
\begin{equation}
    H = - \frac{J}{3} \sum_i^{N-1} (Z_i^{\dagger} Z_{i+1} + Z^{\dagger}_{i+1}Z_i) - h\sum_i^{N} (X_i^{\dagger}  + X_i),
    \label{eq:Potts_H}
\end{equation}
\vspace{-1em}
\begin{equation}
 \quad
Z_i =
\begin{pmatrix}
1 & 0       & 0       \\
0       & \omega & 0       \\
0       & 0       & \omega^2
\end{pmatrix},
\quad\quad
%\quad\text{,}\quad
X_i =
\begin{pmatrix}
0 & 0 & 1 \\
1 & 0 & 0 \\
0 & 1 & 0
\end{pmatrix},
\end{equation}
with $J, h \geq 0$ the interaction strength and the transverse magnetic field respectively, and $\omega = \text{e}^{2 \pi i / 3}$. 
This model has a $\mathbb{Z}_3$ rotational symmetry realized by $\prod_{i=1}^N X_i$, and undergoes a phase transition at the self-dual point $J/3=h$ between a symmetry-broken phase with a threefold degenerate ground state and a paramagnetic phase with a unique ground state~\cite{mong2014parafermionic}.
We realize this model in a spin-1/2 chain of length $L=2N$ by projecting each pair of neighboring spins onto the triplet subspace, onto which the operators $Z_i$ and $X_i$ act.

We characterize the nonstabilizerness in the symmetric ground state for different values of $h$ and $L$ at fixed $J=1$.
The information lattice of the symmetric ground state for $h\ll 1/3$ is shown in the right inset of Fig.~\ref{fig:Potts} with $h= 0$ and $L=12$.
There is an extensive amount of information at small scales $\Omega \sim \mathcal{O}(L)$ and a finite contribution at large scales $\Gamma$ separated by an information gap. 
The same features are also clearly visible in the information per scale as shown in the left inset of Fig.~\ref{fig:Potts}.
The noninteger $\Gamma \approx \log_2(3)$  in the presence of an information gap indicates the long-range nonstabilizerness of the state (see End Matter for further details on this value of $\Gamma$).
For $h\gg 1/3$ the value of $\Gamma$ goes to zero, and local information is present at small scales only, which indicates at most short-range nonstabilizerness.
In Fig.~\ref{fig:Potts} we show both $\Gamma$ and $\Gamma_{\rm folded}$ as functions of $h$ for different system sizes $L$, where curves for different $L$ cross approximately at the phase transition point $h=1/3$.
There is a systematic flow on both sides of the crossings such that $\Gamma$ and $\Gamma_{\rm folded}$ approach either $\log_2(3)$ or $0$ as $L$ increases. 
Away from the critical point, for $h\lesssim 0.2$, the total information at large scales for the folded information lattice is $\Gamma_{\rm folded} \approx \Gamma $, which indicates that long-range nonstabilizerness originates from global correlations.
In the vicinity of the critical point, the information gap closes and nonzero local information is present at all scales.
Thus $\Gamma$ scales with system size and is no longer a witness of long-range nonstabilizerness.
In this case, the folding procedure rearranges local information on the folded information lattice such that there are additional contributions at large scales making $\Gamma_{\rm folded}$ overestimate the global correlations $\gamma_{\rm global}$. 

\textit{Discussion}---We have shown that the information lattice provides a straightforward and direct way to identify nonstabilizerness in any quantum state.
This stems from the fact that local information in a stabilizer state is necessarily an integer.
In localized states, defined by the presence of an information gap, the information lattice allows us to detect long-range nonstabilizerness.
The information gap prevents the redistribution of information between short and large scales under shallow local unitary circuits, making the total information at large scales $\Gamma$ an invariant.
As a consequence, a noninteger $\Gamma$ serves as a witness of long-range nonstabilizerness, since it indicates that a state cannot be transformed into a stabilizer state by any shallow local unitary circuit.
Using a folding procedure we further identify global and edge-to-edge contributions to $\Gamma$.
Our method directly witnesses long-range nonstabilizerness without the need of employing any minimization procedure or a search for a shallow local unitary circuit connecting the state to a stabilizer state.

As an example we have analyzed the long-range nonstabilizerness of the spin-1/2  three-state  Potts model.
At small magnetic field the symmetric ground state of the model has large-scale information $\Gamma \approx \log_2(3)$, indicating long-range nonstabilizerness.
This comes directly from the fact that the ground state is threefold degenerate.
For a large magnetic field, $\Gamma \approx 0$ indicates at most short-range nonstabilizerness.
At the phase transition between these two limits the ground state is no longer localized as the information gap closes and $\Gamma$ cannot be used as a witness of long-range nonstabilizerness.

Since the information lattice is a state property, our approach is independent of any parent Hamiltonian or any protocol used to generate quantum states and aids in the current efforts to characterize states in terms of the scale of their nonstabilizerness.
Witnessing long-range nonstabilizerness from $\Gamma$ in localized states is computationally efficient \cite{artiaco2025universal,aceituno2024thermalization} as compared to brute-force methods involving minimization of nonstabilizerness monotones over all possible local unitary circuits (see End Matter).
The properties discussed in this paper are also potentially accessible through experiments as the density matrix can be reconstructed experimentally through quantum state tomography~\cite{smithey1993measurement, leonhardt1995quantumstate, leibfried1996experimental, lvovsky2009continuous, cramer2010efficient}.

While in this work we focused on one-dimensional systems, a possible future direction is to study nonstabilizerness in localized states in higher dimensions.
Our framework can be readily used via a quasi-$1D$ approach, in which the information lattice of an effective one-dimensional system is obtained by coarse-graining a higher-dimensional system.
More generally, one could explore long-range nonstabilizerness in higher dimensions via a higher-dimensional local decomposition of information~\cite{flor2025highdimensional}.
Another direction is to explore nonstabilizerness in mixed-state phases~\cite{sang2024mixed, sang2025mixed}, to which the information lattice is directly applicable. 
Furthermore, generalizing the characterization of quantum states by the scale of their nonstabilizerness requires constructing a fully local decomposition of nonstabilizerness akin to the information lattice. 
This is, for example, relevant for characterizing critical and ergodic states.

\textit{Acknowledgments}---We thank D. Aceituno Ch\'avez, I. M. Fl\'or and A. Partos for useful discussions. 
This work received funding from the European Research Council (ERC) under the European Union’s Horizon 2020 research and innovation program (Grant Agreement No.~101001902), the Swedish Research Council (VR) through Grant No. 2020-00214, and the Knut and Alice Wallenberg Foundation (KAW) via the project Dynamic Quantum Matter (2019.0068). 
T.~K.~K. acknowledges funding from the Wenner-Gren Foundations. 
The computations were enabled by resources provided by the National Academic Infrastructure for Supercomputing in Sweden (NAISS), partially funded by the Swedish Research Council through grant agreement no. 2022-06725.

\appendix

\section{End Matter}

\subsection{Interpretation of local information in terms of conditional mutual information}
The local information $i_n^{\ell}$~\cite{klein2022time} at position $(n,\ell)$ associated with the subsystem $\mathcal{C}_n^\ell$ is the mutual information shared by the largest lower-scale subsystems $\mathcal{C}_{n-1/2}^{\ell-1}$ and $\mathcal{C}_{n+1/2}^{\ell-1}$ (consisting of sites $[0,\ell-1 ]$ and $[1,\ell]$, respectively) conditioned on the total information available in their overlap, namely the subsystem $\mathcal{C}_n^{\ell-2}$ consisting of sites $[1,\ell-1 ]$.
For example, $i_{4.5}^5$ is the mutual information between subsystems $\mathcal{C}_{4}^{4}$ containing the sites $[2,6]$ and  $\mathcal{C}_{5}^{4}$ containing the sites $ [3,7]$, conditioned on their overlap, which is the subsystem $\mathcal{C}_{4.5}^{3}$ containing sites $[3,6]$.

At scale $\ell = 0$ all subsystems are single-qubit, such that the local information is simply the von Neumann information
\begin{align}
        i_n^{0} \coloneqq I(\mathcal{C}_n^0).
     \label{eq:local_info_ell_0}
\end{align}
At scale $\ell = 1$ the subsystems are two-qubit and can only be partitioned into two single-qubit nonoverlapping subsystems $\mathcal{C}_{n-1/2}^0$ and $\mathcal{C}_{n+1/2}^0$ for any $n$. The local information at any $(n,\ell=1)$ is thus equivalent to mutual information of the lower scale subsystems in the absence of overlapping regions and is given by
\begin{align}
\begin{aligned}
        i_n^1&\coloneqq  i(\mathcal{C}_{n-1/2}^0;\mathcal{C}_{n+1/2}^0)\\& = I(\rho_n^1) - I(\rho_{n-1/2}^0) - I(\rho_{n+1/2}^0).
     \label{eq:local_info_ell_1}
\end{aligned}
\end{align}
At scale  $\ell = 2$, all subsystems $\mathcal{C}_n^2$ consist of three qubits and the largest lower scale subsystems contained within them are $\mathcal{C}_{n-1/2}^1$, $\mathcal{C}_{n+1/2}^1$, each consisting of two qubits, such that they overlap at subsystem $\mathcal{C}_n^0$.
The local information at any $(n,\ell=2)$ is thus the mutual information of the lower-scale two qubit subsystems conditioned on the information available to the single-qubit overlap $\mathcal{C}_{n}^0$  and is given by
\begin{align}
\begin{aligned}
        i_n^2&\coloneqq  i(\mathcal{C}_{n-1/2}^1;\mathcal{C}_{n+1/2}^1\vert \mathcal{C}_{n}^0)\\& = I(\rho_n^2) - I(\rho_{n-1/2}^1) - I(\rho_{n+1/2}^1)+I(\rho^0_n).
     \label{eq:local_info_ell_2}
\end{aligned}
\end{align}
For any $\ell>2$ the definition is the same as in Eq.~\eqref{eq:local_info_ell_2}, where the overlap contains $\ell - 1$ sites, thus
\begin{equation}
\begin{aligned}
i_n^{\ell} \coloneqq 
\begin{cases} 
I(\mathcal{C}_n^\ell), & \ell=0 \\ 
I(\mathcal{C}_{n-1/2}^{\ell-1};\mathcal{C}_{n+1/2}^{\ell-1}\vert \mathcal{C}_{n}^{\ell-2}), & \ell > 1.
\end{cases}
\end{aligned}
\label{joint_meas_cov}
\end{equation}
Written in terms of disjoint subsystems, this definition is equivalent to the conditional mutual information of the leftmost qubit $\mathcal{C}_{n-\ell/2}^{0}$ and the rightmost qubit $\mathcal{C}_{n+\ell/2}^{0}$ conditioned on the intermediate region $\mathcal{C}_{n}^{\ell-2}$, such that $i_n^{\ell} = I(\mathcal{C}_{n-\ell/2}^{0};\mathcal{C}_{n+\ell/2}^{0} | \mathcal{C}_{n}^{\ell-2})$.

\subsection{Von Neumann entropy of stabilizer states}

Following Ref.~\onlinecite{fattal2004entanglement}, the density matrix of an $L$-qubit stabilizer state $\ket{\psi}=\mathcal{U_C}\ket{0}^{\otimes L} $, where $\mathcal{U_C}$ is a Clifford circuit, can be written in the form
\begin{align}
        \rho = \ket{\psi}\bra{\psi}&= 
    \mathcal{U_C}\ket{0}^{\otimes L} \bra{0}^{\otimes L} \mathcal{U_C}^\dagger=
    \mathcal{U_C}\prod_i^L \frac{\mathds{I}+Z_i}{2} \mathcal{U_C}^\dagger= \nonumber \\ 
    &\frac{1}{2^L}\prod_i^L(\mathds{I} + g_i) = 
    \frac{1}{2^L} \sum_{g\in \mathcal{G}} g,
     \label{eq:rho-stab}
\end{align}
where $\mathds{I}$ is the identity matrix, $g$ are the elements of the stabilizer group $\mathcal{G}$ of the state, and $g_i$ are the generators.
Decomposing the chain into two disjoint subsystems, $A$ and $B$, the reduced density matrix of subsystem $A$ can be written as
\begin{align}
    \rho_A=\mathrm{Tr}_B\left(\rho\right) &=  
    \mathrm{Tr}_B\left(  \frac{1}{2^L} \sum_{g\in \mathcal{G}} g \right)=
    \frac{1}{2^L} \sum_{g\in \mathcal{G}} \mathrm{Tr}\left( g_B \right) g_A = \nonumber \\
    &\frac{2^{L_B}}{2^L}\sum_{g_A \in \mathcal{G}_A} g_A=
    \frac{1}{2^{L_A}}\sum_{g_A \in \mathcal{G}_A} g_A,
\end{align}
where $\mathcal{G}_A$ is the stabilizer subgroup for subsystem $A$, $L=L_A + L_B$, and we used the fact that the trace of any Pauli matrix vanishes except in the case of identity, which gives $2^N$.
The von Neumann entropy of subsystem $A$ can be calculated using that
\begin{align}
    &\rho_A^2 = \frac{1}{2^{L_A}} \frac{1}{2^{L_A}} \sum_{g_A \in \mathcal{G}_A} \sum_{\tilde{g}_A \in \mathcal{G}_A} g_A \tilde{g}_A= \nonumber \\
    &\frac{1}{2^{L_A}} \sum_{\tilde{g}_A \in \mathcal{G}_A} \left(\frac{1}{2^{L_A}}  \sum_{g'_A \in \mathcal{G}_A} g_A' \right)=
    \frac{\mathrm{dim} (\mathcal{G}_A)}{2^{L_A}}\rho_A.
\end{align}
This means that the nonzero eigenvalues of the reduced density matrix of a stabilizer state are completely degenerate and equal to $ \lambda = \mathrm{dim}( \mathcal{G}_A )/ 2^{L_A}$, and since $\rho_A$ has unit trace there are exactly $ 2^{N_A} / \mathrm{dim}( \mathcal{G}_A)$ such nonzero eigenvalues.
The von Neumann entropy of a subsystem is then given by
\begin{align}
    &S(\rho_A)=- \mathrm{Tr} \lbrace \rho_A \mathrm{log}\rho_A \rbrace=
    - \sum_\lambda \lambda \mathrm{log}\lambda = \nonumber \\
    &- \frac{2^{L_A}}{\mathrm{dim}( \mathcal{G}_A)}\frac{\mathrm{dim}( \mathcal{G}_A)}{2^{L_A}} \mathrm{log}\left(\frac{\mathrm{dim}( \mathcal{G}_A)}{2^{L_A}} \right)=
    L_A - \vert \mathcal{G}_A \vert,
    \label{EE_subsys_stab}
\end{align}
where $\vert \mathcal{G} \vert$ is the rank of the stabilizer subgroup. 

\subsection{Fine-tuned example of a nonstabilizer state with integer values of local information}
A 4-qubit state with only integer local values on the information lattice is given by $|\Psi\rangle=\sum\limits_{k=1}^{4} \sqrt{\lambda_k} |B_k\rangle_{12} \mathrm{CZ}_{34}|B_k\rangle_{34}$, with $|B_k\rangle$ the four Bell-pair states, $\mathrm{CZ}_{34}$ a controlled $Z$ gate between the third and fourth qubits, and values of $\lambda_k$ given by $(0.0979216462$, $0.0979216462$, $0.0121062183$, $0.7920504894)$. 
These integer values of the local information are not robust under random Clifford evolution, indicating that the state $\ket{\Psi}$ is a nonstabilizer state.

\subsection{The folding procedure}
\begin{figure}[tb]
    \centering
    \includegraphics[width=\columnwidth]{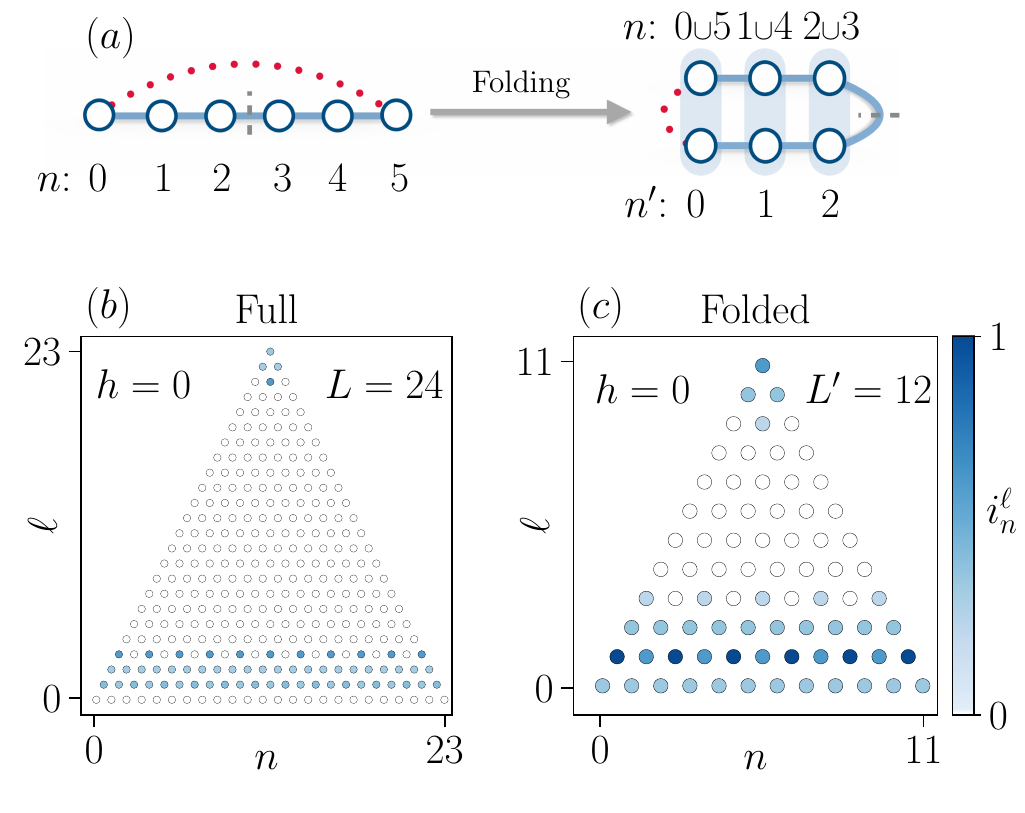}
    \caption{(a) Sketch of the folding procedure used to estimate $\gamma_{\rm global}$ in a chain of length $L=6$. (b) Information lattice of the symmetric ground state of the spin-$1/2$ three-state Potts model at $h=0$ and for system size $L=24$. (c) Folded information lattice for (b).}
    \label{fig:Folding}
\end{figure}
The folding procedure is shown in Fig.~\ref{fig:Folding}(a), where the chain of length $L=6$ is folded across the gray dashed line at its middle point, resulting in a folded chain of length $L'=3$.
The sites in the folded chain are given by the blue ovals, denoted by $n'$, and encompass those sites of the unfolded chain, $n$, belonging to each oval.
The dotted red line sketches an edge-to-edge correlation between sites $n=0$ and $n=5$ in the unfolded chain, which becomes completely local to site $n'=0$ in the folded chain.
Global correlations, on the other hand, do not become local upon folding, and their sum at large scales is unchanged.
An example of the information lattice calculated for the symmetric ground of the of the spin-$1/2$ three-state Potts model from the main text at $h=0$ and $L=24$ is shown in Fig.~\ref{fig:Folding}(b), while its folded counterpart is shown in ~\ref{fig:Folding}(c).

\subsection{$\Gamma$ in symmetry broken states}
Long-range nonstabilizerness may arise in quantum systems when the ground state of the Hamiltonian is degenerate.
This occurs, for example, when there is a spontaneously broken symmetry.
In the symmetric ground state, $\log_2(q)$ bits of information are associated with the relative phases between the different $q$ degenerate states breaking the symmetry.
This results in  $\Gamma = \text{log}_2(q)$, thus $q \neq 2^x$ with $x \in \mathbb{N}$ implies $\Gamma \notin \mathbb{N}$ and long-range nonstabilizerness.
If the degenerate states are product states, the bipartite entanglement entropy $S=\text{log}_2(q)$ across any cut and thus equals $\Gamma$. 
This is, however, not generic, and $S$ does not suffice to extract $\Gamma$.

\subsection{Computational cost}
The cost of calculating the information lattice for a generic state scales exponentially with $L$, the limiting factor being the diagonalization of the density matrices at the scale $\ell\sim L/2$.
In a localized state the nonzero local information resides only at the small and large scales, which can be efficiently calculated \cite{artiaco2025universal,aceituno2024thermalization}.
This is due to the fact that for a pure state the entanglement entropy of the reduced density matrix is equal to that of its complement.
Thus, witnessing long-range nonstabilizerness via $\Gamma$ is computationally less expensive than minimizing a nonstabilizerness monotone over all possible local unitary circuits.

\nocite{zenodo_repo}

%################################
\bibliography{bibliography}

@article{fattal2004entanglement,
    author = "Fattal, David and Cubitt, Toby S. and Yamamoto, Yoshihisa and Bravyi, Sergey and Chuang, Isaac L.",
    title = "{Entanglement in the stabilizer formalism}",
    journal = {arXiv:0406168},
    url={https://arxiv.org/abs/quant-ph/0406168},
}

@article{sharma2025multipartite,
  title = {Multipartite entanglement structures in quantum stabilizer states},
  author = {Sharma, Vaibhav and Mueller, Erich J.},
  journal = {Phys. Rev. A},
  volume = {112},
  issue = {1},
  pages = {012411},
  numpages = {10},
  year = {2025},
  month = {Jul},
  publisher = {American Physical Society},
  doi = {10.1103/1cqt-8rxf},
  url = {https://link.aps.org/doi/10.1103/1cqt-8rxf}
}

@book{zeng2019quantum,
  title={Quantum information meets quantum matter},
  author={Zeng, Bei and Chen, Xie and Zhou, Duan-Lu and Wen, Xiao-Gang and others},
  year={2019},
  publisher={Springer}
}

@article{laflorencie2016quantum,
title = {Quantum entanglement in condensed matter systems},
journal = {Phys. Rep.},
volume = {646},
pages = {1-59},
year = {2016},
issn = {0370-1573},
doi = {https://doi.org/10.1016/j.physrep.2016.06.008},
url = {https://www.sciencedirect.com/science/article/pii/S0370157316301582},
author = {Nicolas Laflorencie},
}

@article{srednicki1993entropy,
  title = {Entropy and area},
  author = {Srednicki, Mark},
  journal = {Phys. Rev. Lett.},
  volume = {71},
  issue = {5},
  pages = {666--669},
  numpages = {0},
  year = {1993},
  month = {Aug},
  publisher = {American Physical Society},
  doi = {10.1103/PhysRevLett.71.666},
  url = {https://link.aps.org/doi/10.1103/PhysRevLett.71.666}
}

@article{hastings2007area,
doi = {10.1088/1742-5468/2007/08/P08024},
url = {https://dx.doi.org/10.1088/1742-5468/2007/08/P08024},
year = {2007},
month = {aug},
publisher = {},
volume = {2007},
number = {08},
pages = {P08024},
author = {M B Hastings},
title = {An area law for one-dimensional quantum systems},
journal = {J. Stat. Mech.}
}

@article{wolf2008area,
  title = {Area Laws in Quantum Systems: Mutual Information and Correlations},
  author = {Wolf, Michael M. and Verstraete, Frank and Hastings, Matthew B. and Cirac, J. Ignacio},
  journal = {Phys. Rev. Lett.},
  volume = {100},
  issue = {7},
  pages = {070502},
  numpages = {4},
  year = {2008},
  month = {Feb},
  publisher = {American Physical Society},
  doi = {10.1103/PhysRevLett.100.070502},
  url = {https://link.aps.org/doi/10.1103/PhysRevLett.100.070502}
}

@article{eisert2010colloquium,
  title = {Colloquium: Area laws for the entanglement entropy},
  author = {Eisert, J. and Cramer, M. and Plenio, M. B.},
  journal = {Rev. Mod. Phys.},
  volume = {82},
  issue = {1},
  pages = {277--306},
  numpages = {0},
  year = {2010},
  month = {Feb},
  publisher = {American Physical Society},
  doi = {10.1103/RevModPhys.82.277},
  url = {https://link.aps.org/doi/10.1103/RevModPhys.82.277}
}

@article{deutsch1991quantum,
  title = {Quantum statistical mechanics in a closed system},
  author = {Deutsch, J. M.},
  journal = {Phys. Rev. A},
  volume = {43},
  number = {4},
  pages = {2046},
  numpages = {0},
  year = {1991},
  publisher = {American Physical Society},
  doi = {10.1103/PhysRevA.43.2046},
}

@article{srednicki1994chaos,
  title = {Chaos and quantum thermalization},
  author = {Srednicki, Mark},
  journal = {Phys. Rev. E},
  volume = {50},
  number = {2},
  pages = {888},
  year = {1994},
  publisher = {American Physical Society},
  doi = {10.1103/PhysRevE.50.888},
}

@article{rigol2008thermalization,
	author = {Rigol, Marcos and Dunjko, Vanja and Olshanii, Maxim},
	date = {2008/04/01},
	doi = {10.1038/nature06838},
	id = {Rigol2008},
	isbn = {1476-4687},
	journal = {Nature},
	number = {7189},
	pages = {854--858},
	title = {Thermalization and its mechanism for generic isolated quantum systems},
	url = {https://doi.org/10.1038/nature06838},
	volume = {452},
	year = {2008}}

@article{kitaev2006topological,
  title = {Topological Entanglement Entropy},
  author = {Kitaev, Alexei and Preskill, John},
  journal = {Phys. Rev. Lett.},
  volume = {96},
  issue = {11},
  pages = {110404},
  numpages = {4},
  year = {2006},
  month = {Mar},
  publisher = {American Physical Society},
  doi = {10.1103/PhysRevLett.96.110404},
  url = {https://link.aps.org/doi/10.1103/PhysRevLett.96.110404}
}

@article{levin2006detecting, 
year = {2006}, 
title = {Detecting Topological Order in a Ground State Wave Function}, 
author = {Levin, Michael and Wen, Xiao-Gang}, 
journal = {Phys. Rev. Lett.}, 
doi = {10.1103/physrevlett.96.110405}, 
pmid = {16605803}, 
pages = {110405}, 
number = {11}, 
volume = {96} 
}

@article{masmendoza2025graphical,
      title={A graphical diagnostic of topological order using {ZX} calculus}, 
      author={Sergi Mas-Mendoza and Richard D. P. East and Michele Filippone and Adolfo G. Grushin},
      journal={arXiv:2509.12355},
      url={https://arxiv.org/abs/2509.12355}, 
}

@article{hamma2005ground,
title = {Ground state entanglement and geometric entropy in the {Kitaev} model},
journal = {Phys. Lett. A},
volume = {337},
number = {1},
pages = {22-28},
year = {2005},
issn = {0375-9601},
doi = {https://doi.org/10.1016/j.physleta.2005.01.060},
url = {https://www.sciencedirect.com/science/article/pii/S0375960105001544},
author = {Alioscia Hamma and Radu Ionicioiu and Paolo Zanardi},
}

@article{leone2022stabilizer,
  title = {Stabilizer {R\'enyi} Entropy},
  author = {Leone, Lorenzo and Oliviero, Salvatore F. E. and Hamma, Alioscia},
  journal = {Phys. Rev. Lett.},
  volume = {128},
  issue = {5},
  pages = {050402},
  numpages = {5},
  year = {2022},
  month = {Feb},
  publisher = {American Physical Society},
  doi = {10.1103/PhysRevLett.128.050402},
  url = {https://link.aps.org/doi/10.1103/PhysRevLett.128.050402}
}

@article{howard2017application,
  title = {Application of a Resource Theory for Magic States to Fault-Tolerant Quantum Computing},
  author = {Howard, Mark and Campbell, Earl},
  journal = {Phys. Rev. Lett.},
  volume = {118},
  issue = {9},
  pages = {090501},
  numpages = {6},
  year = {2017},
  month = {Mar},
  publisher = {American Physical Society},
  doi = {10.1103/PhysRevLett.118.090501},
  url = {https://link.aps.org/doi/10.1103/PhysRevLett.118.090501}
}

@article{bravyi2019simulation,
  doi = {10.22331/q-2019-09-02-181},
  url = {https://doi.org/10.22331/q-2019-09-02-181},
  title = {Simulation of quantum circuits by low-rank stabilizer decompositions},
  author = {Bravyi, Sergey and Browne, Dan and Calpin, Padraic and Campbell, Earl and Gosset, David and Howard, Mark},
  journal = {{Quantum}},
  issn = {2521-327X},
  publisher = {{Verein zur F{\"{o}}rderung des Open Access Publizierens in den Quantenwissenschaften}},
  volume = {3},
  pages = {181},
  month = sep,
  year = {2019}
}

@article{emerson2013theresource,
    author = "Emerson, Joseph and Gottesman, Daniel and Mousavian, Seyed Ali Hamed and Veitch, Victor",
    title = "{The resource theory of stabilizer quantum computation}",
    doi = "10.1088/1367-2630/16/1/013009",
    journal = "New J. Phys.",
    volume = "16",
    number = "1",
    pages = "013009",
    year = "2014"
}

@article{jiang2023lower,
  title = {Lower Bound for the {$T$} Count Via Unitary Stabilizer Nullity},
  author = {Jiang, Jiaqing and Wang, Xin},
  journal = {Phys. Rev. Appl.},
  volume = {19},
  issue = {3},
  pages = {034052},
  numpages = {19},
  year = {2023},
  month = {Mar},
  publisher = {American Physical Society},
  doi = {10.1103/PhysRevApplied.19.034052},
  url = {https://link.aps.org/doi/10.1103/PhysRevApplied.19.034052}
}

@article{dowling2025magic,
  title = {Magic Resources of the {H}eisenberg Picture},
  author = {Dowling, Neil and Kos, Pavel and Turkeshi, Xhek},
  journal = {Phys. Rev. Lett.},
  volume = {135},
  issue = {5},
  pages = {050401},
  numpages = {10},
  year = {2025},
  month = {Jul},
  publisher = {American Physical Society},
  doi = {10.1103/p7xt-s9nz},
  url = {https://link.aps.org/doi/10.1103/p7xt-s9nz}
}

@article{dowling2025bridging,
  title = {Bridging Entanglement and Magic Resources within Operator Space},
  author = {Dowling, Neil and Modi, Kavan and White, Gregory A. L.},
  journal = {Phys. Rev. Lett.},
  volume = {135},
  issue = {16},
  pages = {160201},
  numpages = {9},
  year = {2025},
  month = {Oct},
  publisher = {American Physical Society},
  doi = {10.1103/c7k1-xcwy},
  url = {https://link.aps.org/doi/10.1103/c7k1-xcwy}
}

@article{tirrito2024quantifying,
  title = {Quantifying nonstabilizerness through entanglement spectrum flatness},
  author = {Tirrito, Emanuele and Tarabunga, Poetri Sonya and Lami, Gugliemo and Chanda, Titas and Leone, Lorenzo and Oliviero, Salvatore F. E. and Dalmonte, Marcello and Collura, Mario and Hamma, Alioscia},
  journal = {Phys. Rev. A},
  volume = {109},
  issue = {4},
  pages = {L040401},
  numpages = {6},
  year = {2024},
  month = {Apr},
  publisher = {American Physical Society},
  doi = {10.1103/PhysRevA.109.L040401},
  url = {https://link.aps.org/doi/10.1103/PhysRevA.109.L040401}
}

@article{paviglianiti2025estimating,
  title = {Estimating Nonstabilizerness Dynamics Without Simulating It},
  author = {Paviglianiti, Alessio and Lami, Guglielmo and Collura, Mario and Silva, Alessandro},
  journal = {PRX Quantum},
  volume = {6},
  issue = {3},
  pages = {030320},
  numpages = {17},
  year = {2025},
  month = {Aug},
  publisher = {American Physical Society},
  doi = {10.1103/msm2-vmg7},
  url = {https://link.aps.org/doi/10.1103/msm2-vmg7}
}

@article{gottesman1998heisenberg,
      title={The {Heisenberg} Representation of Quantum Computers}, 
      author={D. Gottesman},
      year={1998},
      journal={arXiv:9807006},
      url={https://arxiv.org/abs/quant-ph/9807006}
}

@article{knill1996nonbinary,
      title={Non-binary Unitary Error Bases and Quantum Codes}, 
      author={E. Knill},
      journal={arXiv:9608048},
      url={https://arxiv.org/abs/quant-ph/9608048} 
}

@article{gottesman1998theory,
  title = {Theory of fault-tolerant quantum computation},
  author = {Gottesman, Daniel},
  journal = {Phys. Rev. A},
  volume = {57},
  issue = {1},
  pages = {127--137},
  numpages = {0},
  year = {1998},
  month = {Jan},
  publisher = {American Physical Society},
  doi = {10.1103/PhysRevA.57.127},
  url = {https://link.aps.org/doi/10.1103/PhysRevA.57.127}
}

@article{gottesman1997stabilizer,
      title={Stabilizer Codes and Quantum Error Correction}, 
      author={Daniel Gottesman},
      journal={arXiv:9705052},
      url={https://arxiv.org/abs/quant-ph/9705052}, 
}

@article{bravyi2005universal,
  title = {Universal quantum computation with ideal {Clifford} gates and noisy ancillas},
  author = {Bravyi, Sergey and Kitaev, Alexei},
  journal = {Phys. Rev. A},
  volume = {71},
  issue = {2},
  pages = {022316},
  numpages = {14},
  year = {2005},
  month = {Feb},
  publisher = {American Physical Society},
  doi = {10.1103/PhysRevA.71.022316},
  url = {https://link.aps.org/doi/10.1103/PhysRevA.71.022316}
}

@article{fowler2012surface,
  title = {Surface codes: Towards practical large-scale quantum computation},
  author = {Fowler, Austin G. and Mariantoni, Matteo and Martinis, John M. and Cleland, Andrew N.},
  journal = {Phys. Rev. A},
  volume = {86},
  issue = {3},
  pages = {032324},
  numpages = {48},
  year = {2012},
  month = {Sep},
  publisher = {American Physical Society},
  doi = {10.1103/PhysRevA.86.032324},
  url = {https://link.aps.org/doi/10.1103/PhysRevA.86.032324}
}

@article{turkeshi2025pauli,
  title = {Pauli spectrum and nonstabilizerness of typical quantum many-body states},
  author = {Turkeshi, Xhek and Dymarsky, Anatoly and Sierant, Piotr},
  journal = {Phys. Rev. B},
  volume = {111},
  issue = {5},
  pages = {054301},
  numpages = {12},
  year = {2025},
  month = {Feb},
  publisher = {American Physical Society},
  doi = {10.1103/PhysRevB.111.054301},
  url = {https://link.aps.org/doi/10.1103/PhysRevB.111.054301}
}

@article{oliviero2022magicstate,
  title = {Magic-state resource theory for the ground state of the transverse-field {Ising} model},
  author = {Oliviero, Salvatore F. E. and Leone, Lorenzo and Hamma, Alioscia},
  journal = {Phys. Rev. A},
  volume = {106},
  issue = {4},
  pages = {042426},
  numpages = {6},
  year = {2022},
  month = {Oct},
  publisher = {American Physical Society},
  doi = {10.1103/PhysRevA.106.042426},
  url = {https://link.aps.org/doi/10.1103/PhysRevA.106.042426}
}

@article{passarelli2024nonstabilizerness,
  title = {Nonstabilizerness of permutationally invariant systems},
  author = {Passarelli, G. and Fazio, R. and Lucignano, P.},
  journal = {Phys. Rev. A},
  volume = {110},
  issue = {2},
  pages = {022436},
  numpages = {9},
  year = {2024},
  month = {Aug},
  publisher = {American Physical Society},
  doi = {10.1103/PhysRevA.110.022436},
  url = {https://link.aps.org/doi/10.1103/PhysRevA.110.022436}
}

@article{passarelli2025chaos,
  doi = {10.22331/q-2025-03-05-1653},
  url = {https://doi.org/10.22331/q-2025-03-05-1653},
  title = {Chaos and magic in the dissipative quantum kicked top},
  author = {Passarelli, Gianluca and Lucignano, Procolo and Rossini, Davide and Russomanno, Angelo},
  journal = {{Quantum}},
  issn = {2521-327X},
  publisher = {{Verein zur F{\"{o}}rderung des Open Access Publizierens in den Quantenwissenschaften}},
  volume = {9},
  pages = {1653},
  month = mar,
  year = {2025}
}

@article{rattasaco2023stabilizer,
  title = {Stabilizer entropy dynamics after a quantum quench},
  author = {Rattacaso, Davide and Leone, Lorenzo and Oliviero, Salvatore F. E. and Hamma, Alioscia},
  journal = {Phys. Rev. A},
  volume = {108},
  issue = {4},
  pages = {042407},
  numpages = {9},
  year = {2023},
  month = {Oct},
  publisher = {American Physical Society},
  doi = {10.1103/PhysRevA.108.042407},
  url = {https://link.aps.org/doi/10.1103/PhysRevA.108.042407}
}

@article{tirrito2025anticoncentration,
      title={Anticoncentration and nonstabilizerness spreading under ergodic quantum dynamics}, 
      author={Emanuele Tirrito and Xhek Turkeshi and Piotr Sierant},
      year={2024},
      journal={arXiv:2412.10229},
      url={https://arxiv.org/abs/2412.10229}, 
}

@article{tirrito2025magic,
      title={Magic phase transitions in monitored gaussian fermions}, 
      author={Emanuele Tirrito and Luca Lumia and Alessio Paviglianiti and Guglielmo Lami and Alessandro Silva and Xhek Turkeshi and Mario Collura},
      year={2025},
      journal={arXiv:2507.07179},
      url={https://arxiv.org/abs/2507.07179}, 
}

@article{russomanno2025nonstabilizerness,
  title = {Nonstabilizerness in the unitary and monitored quantum dynamics of {XXZ}-staggered and {Sachdev}-{Ye}-{Kitaev} models},
  author = {Russomanno, Angelo and Passarelli, Gianluca and Rossini, Davide and Lucignano, Procolo},
  journal = {Phys. Rev. B},
  volume = {112},
  issue = {6},
  pages = {064312},
  numpages = {11},
  year = {2025},
  month = {Aug},
  publisher = {American Physical Society},
  doi = {10.1103/njgn-fksh},
  url = {https://link.aps.org/doi/10.1103/njgn-fksh}
}

@article{odavic2025stabilizer,
  title = {Stabilizer entropy in nonintegrable quantum evolutions},
  author = {Odavi\ifmmode \acute{c}\else \'{c}\fi{}, J. and Viscardi, M. and Hamma, A.},
  journal = {Phys. Rev. B},
  volume = {112},
  issue = {10},
  pages = {104301},
  numpages = {14},
  year = {2025},
  month = {Sep},
  publisher = {American Physical Society},
  doi = {10.1103/y9r6-dx7p},
  url = {https://link.aps.org/doi/10.1103/y9r6-dx7p}
}

@article{viscardi2025interplay,
	title = {Interplay of entanglement structures and stabilizer entropy in spin models},
	pages = {012},
	author = {Viscardi, Michele and Dalmonte, Marcello and Hamma, Alioscia and Tirrito, Emanuele},
	journal = {SciPost Phys. Core},
	volume = {9},
	year = {2026},
	publisher = {SciPost},
	doi = {10.21468/SciPostPhysCore.9.1.012},
	url = {https://scipost.org/10.21468/SciPostPhysCore.9.1.012}
}

@article{collura2025nonstabilizerness,
  doi = {10.22331/q-2026-03-23-2036},
  url = {https://doi.org/10.22331/q-2026-03-23-2036},
  title = {The non-stabilizerness of fermionic {G}aussian states},
  author = {Collura, Mario and Nardis, Jacopo De and Alba, Vincenzo and Lami, Guglielmo},
  journal = {{Quantum}},
  issn = {2521-327X},
  publisher = {{Verein zur F{\"{o}}rderung des Open Access Publizierens in den Quantenwissenschaften}},
  volume = {10},
  pages = {2036},
  month = mar,
  year = {2026}
}

@article{sarkis2025aremolecules,
      title={Are Molecules Magical? {Non-Stabilizerness} in Molecular Bonding}, 
      author={Matthieu Sarkis and Alexandre Tkatchenko},
      journal={arXiv:2504.06673},
      url={https://arxiv.org/abs/2504.06673}, 
}

@article{turkeshi2025magic,
  title={Magic spreading in random quantum circuits},
  author={Turkeshi, Xhek and Tirrito, Emanuele and Sierant, Piotr},
  journal={Nat. Commun.},
  volume={16},
  number={1},
  pages={2575},
  year={2025},
  publisher={Nature Publishing Group UK London},
  url={https://www.nature.com/articles/s41467-025-57704-x}, 
}

@article{falcao2025magic,
  title={Magic dynamics in many-body localized systems},
  author={Falc{\~a}o, Pedro R Nic{\'a}cio and Sierant},
  journal={arXiv:2503.07468},
  url={https://arxiv.org/abs/2503.07468},
}

@article{hernandez2025non,
  title={Non-stabilizerness in quantum-enhanced metrological protocols},
  author={Hern{\'a}ndez-Yanes, Tanaus{\'u} and Sierant, Piotr and Zakrzewski, Jakub and P{\l}odzie{\'n}, Marcin},
  journal={arXiv:2510.01380},
  url = {https://arxiv.org/abs/2510.01380},
}

@article{sarkar2020characterization,
  title={Characterization of an operational quantum resource in a critical many-body system},
  author={Sarkar, Saubhik and Mukhopadhyay, Chiranjib and Bayat, Abolfazl},
  journal={New J. Phys.},
  volume={22},
  number={8},
  pages={083077},
  year={2020},
  publisher={IOP Publishing},
  url={https://iopscience.iop.org/article/10.1088/1367-2630/aba919},
}

@article{santra2025complexity,
      title={Complexity transitions in chaotic quantum systems: {Nonstabilizerness}, entanglement, and fractal dimension in {SYK} and random matrix models}, 
      author={Gopal Chandra Santra and Alex Windey and Soumik Bandyopadhyay and Andrea Legramandi and Philipp Hauke},
      journal={arXiv:2505.09707},
      url={https://arxiv.org/abs/2505.09707}, 
}

@article{haug2025probing,
  doi = {10.22331/q-2025-07-21-1801},
  url = {https://doi.org/10.22331/q-2025-07-21-1801},
  title = {Probing quantum complexity via universal saturation of stabilizer entropies},
  author = {Haug, Tobias and Aolita, Leandro and Kim, M.S.},
  journal = {{Quantum}},
  issn = {2521-327X},
  publisher = {{Verein zur F{\"{o}}rderung des Open Access Publizierens in den Quantenwissenschaften}},
  volume = {9},
  pages = {1801},
  month = jul,
  year = {2025}
}

@article{sierant2025fermionic,
  title = {Fermionic Magic Resources of Quantum Many-Body Systems},
  author = {Sierant, Piotr and Stornati, Paolo and Turkeshi, Xhek},
  journal = {PRX Quantum},
  volume = {7},
  issue = {1},
  pages = {010302},
  numpages = {41},
  year = {2026},
  month = {Jan},
  publisher = {American Physical Society},
  doi = {10.1103/3yx4-1j27},
  url = {https://link.aps.org/doi/10.1103/3yx4-1j27}
}

@article{terhal2002classical,
  title = {Classical simulation of noninteracting-fermion quantum circuits},
  author = {Terhal, Barbara M. and DiVincenzo, David P.},
  journal = {Phys. Rev. A},
  volume = {65},
  issue = {3},
  pages = {032325},
  numpages = {10},
  year = {2002},
  month = {Mar},
  publisher = {American Physical Society},
  doi = {10.1103/PhysRevA.65.032325},
  url = {https://link.aps.org/doi/10.1103/PhysRevA.65.032325}
}

@inproceedings{valiant2001quantum,
author = {Valiant, Leslie G.},
title = {Quantum computers that can be simulated classically in polynomial time},
year = {2001},
isbn = {1581133499},
publisher = {Association for Computing Machinery},
address = {New York, NY, USA},
url = {https://doi.org/10.1145/380752.380785},
doi = {10.1145/380752.380785},
booktitle = {Proceedings of the Thirty-Third Annual ACM Symposium on Theory of Computing},
pages = {114–123},
numpages = {10},
location = {Hersonissos, Greece},
series = {STOC '01}
}

@article{qian2025quantum,
  title = {Quantum nonlocal nonstabilizerness},
  author = {Qian, Dongheng and Wang, Jing},
  journal = {Phys. Rev. A},
  volume = {111},
  issue = {5},
  pages = {052443},
  numpages = {9},
  year = {2025},
  month = {May},
  publisher = {American Physical Society},
  doi = {10.1103/PhysRevA.111.052443},
  url = {https://link.aps.org/doi/10.1103/PhysRevA.111.052443}
}

@article{cao2025nontrivial,
  title = {Non-trivial area operators require non-local magic},
  author = {Cao, ChunJun},
  journal = {JHEP},
  volume  = {11},
  year    = {2024},
  pages   = {105  (2024)},
  doi     = {10.1007/JHEP11(2024)105},
  url = {https://doi.org/10.1007/JHEP11(2024)105},
}

@article{huang2025nonstabilizerness,
  title = {Nonstabilizerness entanglement entropy: A measure of hardness in the classical simulation of quantum many-body systems with tensor network states},
  author = {Huang, Jiale and Qian, Xiangjian and Qin, Mingpu},
  journal = {Phys. Rev. A},
  volume = {112},
  issue = {1},
  pages = {012425},
  numpages = {11},
  year = {2025},
  month = {Jul},
  publisher = {American Physical Society},
  doi = {10.1103/gxdn-zwrw},
  url = {https://link.aps.org/doi/10.1103/gxdn-zwrw}
}

@article{cao2025gravitational,
  title = {Gravitational Backreaction is Magical},
  author = {Cao, ChunJun and Cheng, Gong and Hamma, Alioscia and Leone, Lorenzo and Munizzi, William and Oliviero, Savatore F.E.},
  journal = {PRX Quantum},
  volume = {6},
  issue = {4},
  pages = {040375},
  numpages = {39},
  year = {2025},
  month = {Dec},
  publisher = {American Physical Society},
  doi = {10.1103/z3vr-w5c5},
  url = {https://link.aps.org/doi/10.1103/z3vr-w5c5}
}

@article{korbany2025longrange,
  title = {Long-Range Nonstabilizerness and Phases of Matter},
  author = {Korbany, David Aram and Gullans, Michael J. and Piroli, Lorenzo},
  journal = {Phys. Rev. Lett.},
  volume = {135},
  issue = {16},
  pages = {160404},
  numpages = {7},
  year = {2025},
  month = {Oct},
  publisher = {American Physical Society},
  doi = {10.1103/1hlj-h6t9},
  url = {https://link.aps.org/doi/10.1103/1hlj-h6t9}
}

@article{white2021conformal,
  title = {Conformal field theories are magical},
  author = {White, Christopher David and Cao, ChunJun and Swingle, Brian},
  journal = {Phys. Rev. B},
  volume = {103},
  issue = {7},
  pages = {075145},
  numpages = {18},
  year = {2021},
  month = {Feb},
  publisher = {American Physical Society},
  doi = {10.1103/PhysRevB.103.075145},
  url = {https://link.aps.org/doi/10.1103/PhysRevB.103.075145}
}

@article{ellison2021symmetry,
  doi = {10.22331/q-2021-12-28-612},
  url = {https://doi.org/10.22331/q-2021-12-28-612},
  title = {Symmetry-protected sign problem and magic in quantum phases of matter},
  author = {Ellison, Tyler D. and Kato, Kohtaro and Liu, Zi-Wen and Hsieh, Timothy H.},
  journal = {{Quantum}},
  issn = {2521-327X},
  publisher = {{Verein zur F{\"{o}}rderung des Open Access Publizierens in den Quantenwissenschaften}},
  volume = {5},
  pages = {612},
  month = dec,
  year = {2021}
}

@article{tarabunga2025efficient,
	title={{Efficient mutual magic and magic capacity with matrix product states}},
	author={Poetri Sonya Tarabunga and Tobias Haug},
	journal={SciPost Phys.},
	volume={19},
	pages={085},
	year={2025},
	publisher={SciPost},
	doi={10.21468/SciPostPhys.19.4.085},
	url={https://scipost.org/10.21468/SciPostPhys.19.4.085},
}

@article{timsina2025robustness,
      title={Robustness of {Magic} in the quantum {Ising} chain via {Quantum} {Monte} {Carlo} tomography}, 
      author={Hari Timsina and Yi-Ming Ding and Emanuele Tirrito and Poetri Sonya Tarabunga and Bin-Bin Mao and Mario Collura and Zheng Yan and Marcello Dalmonte},
      journal={arXiv:2507.12902},
      url={https://arxiv.org/abs/2507.12902}, 
}

@article{chen2010local,
  title = {Local unitary transformation, long-range quantum entanglement, wave function renormalization, and topological order},
  author = {Chen, Xie and Gu, Zheng-Cheng and Wen, Xiao-Gang},
  journal = {Phys. Rev. B},
  volume = {82},
  issue = {15},
  pages = {155138},
  numpages = {28},
  year = {2010},
  month = {Oct},
  publisher = {American Physical Society},
  doi = {10.1103/PhysRevB.82.155138},
  url = {https://link.aps.org/doi/10.1103/PhysRevB.82.155138}
}

@article{klein2022time,
  title={Time-evolution of local information: thermalization dynamics of local observables},
  author={Klein Kvorning, Thomas and Herviou, Lo{\"\i}c and Bardarson, Jens H},
  journal={SciPost Phys.},
  volume={13},
  number={4},
  pages={080},
  year={2022},
doi={10.21468/SciPostPhys.13.4.080}
}

@article{artiaco2024efficient,
  title = {Efficient Large-Scale Many-Body Quantum Dynamics via Local-Information Time Evolution},
  author = {Artiaco, Claudia and Fleckenstein, Christoph and Aceituno Ch\'avez, David and Klein Kvorning, Thomas and Bardarson, Jens H.},
  journal = {PRX Quantum},
  volume = {5},
  issue = {2},
  pages = {020352},
  numpages = {28},
  year = {2024},
  month = {Jun},
  publisher = {American Physical Society},
  doi = {10.1103/PRXQuantum.5.020352},
  url = {https://link.aps.org/doi/10.1103/PRXQuantum.5.020352}
}

@article{harkins2025nanoscale,
    author = {Kieren Harkins  and Christoph Fleckenstein  and Noella D’Souza  and Paul M. Schindler  and David Marchiori  and Claudia Artiaco  and Quentin Reynard-Feytis  and Ushoshi Basumallick  and William Beatrez  and Arjun Pillai  and Matthias Hagn  and Aniruddha Nayak  and Samantha Breuer  and Xudong Lv  and Maxwell McAllister  and Paul Reshetikhin  and Emanuel Druga  and Marin Bukov  and Ashok Ajoy },
    title = {Nanoscale engineering and dynamic stabilization of mesoscopic spin textures},
    journal = {Sci. Adv.},
    volume = {11},
    number = {13},
    pages = {eadn9021},
    year = {2025},
    doi = {10.1126/sciadv.adn9021},
    URL = {https://www.science.org/doi/abs/10.1126/sciadv.adn9021}
}

@article{artiaco2025universal,
  title = {Universal Characterization of Quantum Many-Body States through Local Information},
  author = {Artiaco, Claudia and Klein Kvorning, Thomas and Aceituno Ch\'avez, David and Herviou, Lo\"{\i}c and Bardarson, Jens H.},
  journal = {Phys. Rev. Lett.},
  volume = {134},
  issue = {19},
  pages = {190401},
  numpages = {8},
  year = {2025},
  month = {May},
  publisher = {American Physical Society},
  doi = {10.1103/PhysRevLett.134.190401},
  url = {https://link.aps.org/doi/10.1103/PhysRevLett.134.190401}
}

@article{bauer2025local,
  title = {Local information flow in quantum quench dynamics},
  author = {Bauer, Nicolas P. and Trauzettel, Bj\"orn and Klein Kvorning, Thomas and Bardarson, Jens H. and Artiaco, Claudia},
  journal = {Phys. Rev. A},
  volume = {112},
  issue = {2},
  pages = {022221},
  numpages = {14},
  year = {2025},
  month = {Aug},
  publisher = {American Physical Society},
  doi = {10.1103/v7gb-5gq8},
  url = {https://link.aps.org/doi/10.1103/v7gb-5gq8}
}

@article{artiaco2025out,
      title={Out-of-Equilibrium Dynamics in a {U}(1) Lattice Gauge Theory via Local Information Flows: Scattering and String Breaking}, 
      author={Artiaco, Claudia and Barata, Jo{\~a}o and Rico, Enrique},
      journal={arXiv:2510.16101},
      url={https://arxiv.org/abs/2510.16101},
}

@article{nava2026information,
  title = {Information transport and transport-induced entanglement in open fermion chains},
  author = {Nava, Andrea and Artiaco, Claudia and Gefen, Yuval and Gornyi, Igor and Tsitsishvili, Mikheil and Zazunov, Alex and Egger, Reinhold},
  journal = {Phys. Rev. Res.},
  volume = {8},
  issue = {3},
  pages = {033273},
  numpages = {28},
  year = {2026},
  month = {Sep},
  publisher = {American Physical Society},
  doi = {10.1103/5dks-395l},
  url = {https://link.aps.org/doi/10.1103/5dks-395l}
}

@article{barata2025hadronic,
  title={Hadronic scattering in (1+ 1){D} {SU}(2) lattice gauge theory from tensor networks},
  author={Barata, Jo{\~a}o and Hormaza, Juan and Kang, Zhong-Bo and Qian, Wenyang},
  journal = {arXiv:2511.00154},
  url = {https://arxiv.org/abs/2511.00154}
}

@article{flor2025highdimensional,
      title={Higher-Dimensional Information Lattice: Quantum State Characterization through Inclusion-Exclusion Local Information}, 
      author={Ian Matthias Flór and Claudia Artiaco and Thomas Klein Kvorning and Jens H. Bardarson},
      journal={arXiv:2512.20793},
      url={https://arxiv.org/abs/2512.20793}, 
}

@article{tarabunga2024critical,
  doi = {10.22331/q-2024-07-17-1413},
  url = {https://doi.org/10.22331/q-2024-07-17-1413},
  title = {Critical behaviors of non-stabilizerness in quantum spin chains},
  author = {Tarabunga, Poetri Sonya},
  journal = {{Quantum}},
  issn = {2521-327X},
  publisher = {{Verein zur F{\"{o}}rderung des Open Access Publizierens in den Quantenwissenschaften}},
  volume = {8},
  pages = {1413},
  month = jul,
  year = {2024}
}

@article{tarabunga2023manybody,
  title = {Many-Body Magic Via {Pauli}-{Markov} Chains---From Criticality to Gauge Theories},
  author = {Tarabunga, Poetri Sonya and Tirrito, Emanuele and Chanda, Titas and Dalmonte, Marcello},
  journal = {PRX Quantum},
  volume = {4},
  issue = {4},
  pages = {040317},
  numpages = {19},
  year = {2023},
  month = {Oct},
  publisher = {American Physical Society},
  doi = {10.1103/PRXQuantum.4.040317},
  url = {https://link.aps.org/doi/10.1103/PRXQuantum.4.040317}
}

@article{wu1982thepotts,
  title = {The {Potts} model},
  author = {Wu, F. Y.},
  journal = {Rev. Mod. Phys.},
  volume = {54},
  issue = {1},
  pages = {235--268},
  numpages = {0},
  year = {1982},
  month = {Jan},
  publisher = {American Physical Society},
  doi = {10.1103/RevModPhys.54.235},
  url = {https://link.aps.org/doi/10.1103/RevModPhys.54.235}
}

@article{mong2014parafermionic,
doi = {10.1088/1751-8113/47/45/452001},
url = {https://doi.org/10.1088/1751-8113/47/45/452001},
year = {2014},
month = {oct},
publisher = {IOP Publishing},
volume = {47},
number = {45},
pages = {452001},
author = {Mong, Roger S K and Clarke, David J and Alicea, Jason and Lindner, Netanel H and Fendley, Paul},
title = {Parafermionic conformal field theory on the lattice},
journal = {J. Phys. A},
}

@article{smithey1993measurement,
  title = {Measurement of the Wigner distribution and the density matrix of a light mode using optical homodyne tomography: {Application} to squeezed states and the vacuum},
  author = {Smithey, D. T. and Beck, M. and Raymer, M. G. and Faridani, A.},
  journal = {Phys. Rev. Lett.},
  volume = {70},
  issue = {9},
  pages = {1244--1247},
  numpages = {0},
  year = {1993},
  month = {Mar},
  publisher = {American Physical Society},
  doi = {10.1103/PhysRevLett.70.1244},
  url = {https://link.aps.org/doi/10.1103/PhysRevLett.70.1244}
}

@article{leibfried1996experimental,
  title = {Experimental Determination of the Motional Quantum State of a Trapped Atom},
  author = {Leibfried, D. and Meekhof, D. M. and King, B. E. and Monroe, C. and Itano, W. M. and Wineland, D. J.},
  journal = {Phys. Rev. Lett.},
  volume = {77},
  issue = {21},
  pages = {4281--4285},
  numpages = {0},
  year = {1996},
  month = {Nov},
  publisher = {American Physical Society},
  doi = {10.1103/PhysRevLett.77.4281},
  url = {https://link.aps.org/doi/10.1103/PhysRevLett.77.4281}
}

@article{leonhardt1995quantumstate,
  title = {Quantum-State Tomography and Discrete {Wigner} Function},
  author = {Leonhardt, Ulf},
  journal = {Phys. Rev. Lett.},
  volume = {74},
  issue = {21},
  pages = {4101--4105},
  numpages = {0},
  year = {1995},
  month = {May},
  publisher = {American Physical Society},
  doi = {10.1103/PhysRevLett.74.4101},
  url = {https://link.aps.org/doi/10.1103/PhysRevLett.74.4101}
}

@article{cramer2010efficient,
author={Cramer, Marcus
and Plenio, Martin B.
and Flammia, Steven T.
and Somma, Rolando
and Gross, David
and Bartlett, Stephen D.
and Landon-Cardinal, Olivier
and Poulin, David
and Liu, Yi-Kai},
title={Efficient quantum state tomography},
journal={Nat. Commun.},
year={2010},
month={Dec},
day={21},
volume={1},
number={1},
pages={149},
issn={2041-1723},
doi={10.1038/ncomms1147},
url={https://doi.org/10.1038/ncomms1147}
}

@article{lvovsky2009continuous,
  title = {Continuous-variable optical quantum-state tomography},
  author = {Lvovsky, A. I. and Raymer, M. G.},
  journal = {Rev. Mod. Phys.},
  volume = {81},
  issue = {1},
  pages = {299--332},
  numpages = {0},
  year = {2009},
  month = {Mar},
  publisher = {American Physical Society},
  doi = {10.1103/RevModPhys.81.299},
  url = {https://link.aps.org/doi/10.1103/RevModPhys.81.299}
}

@article{fux2025disentangling,
  title = {Disentangling Magic States with Classically Simulable Quantum Circuits},
  author = {Fux, Gerald E. and B{\'e}ri, Benjamin and Fazio, Rosario and Tirrito, Emanuele},
  journal = {Phys. Rev. Lett.},
  volume = {135},
  issue = {26},
  pages = {260605},
  numpages = {8},
  year = {2025},
  month = {Dec},
  publisher = {American Physical Society},
  doi = {10.1103/ggp1-byj1},
  url = {https://link.aps.org/doi/10.1103/ggp1-byj1}
}

@article{bejan2024dynamical, 
year = {2024}, 
title = {{Dynamical Magic Transitions in Monitored Clifford+T Circuits}}, 
author = {Bejan, Mircea and McLauchlan, Campbell and Béri, Benjamin}, 
journal = {PRX Quantum}, 
doi = {10.1103/prxquantum.5.030332}, 
pages = {030332}, 
number = {3}, 
volume = {5}, 
}

@article{Sang2024mixed,
  title = {Mixed-State Quantum Phases: Renormalization and Quantum Error Correction},
  author = {Sang, Shengqi and Zou, Yijian and Hsieh, Timothy H.},
  journal = {Phys. Rev. X},
  volume = {14},
  issue = {3},
  pages = {031044},
  numpages = {24},
  year = {2024},
  month = {Sep},
  publisher = {American Physical Society},
  doi = {10.1103/PhysRevX.14.031044},
  url = {https://link.aps.org/doi/10.1103/PhysRevX.14.031044}
}

@article{sang2025mixed,
      title={Mixed-state phases from local reversibility}, 
      author={Shengqi Sang and Leonardo A. Lessa and Roger S. K. Mong and Tarun Grover and Chong Wang and Timothy H. Hsieh},
      journal={arXiv:2507.02292},
      url={https://arxiv.org/abs/2507.02292}, 
}

@article{schollwock2011thedensitymatrix,
title = {The density-matrix renormalization group in the age of matrix product states},
journal = {Ann. Phys.},
volume = {326},
number = {1},
pages = {96-192},
year = {2011},
issn = {0003-4916},
doi = {https://doi.org/10.1016/j.aop.2010.09.012},
url = {https://www.sciencedirect.com/science/article/pii/S0003491610001752},
author = {Ulrich Schollwöck},
}

@article{cirac2021matrix,
  title = {Matrix product states and projected entangled pair states: Concepts, symmetries, theorems},
  author = {Cirac, J. Ignacio and P\'erez-Garc\'{\i}a, David and Schuch, Norbert and Verstraete, Frank},
  journal = {Rev. Mod. Phys.},
  volume = {93},
  issue = {4},
  pages = {045003},
  numpages = {65},
  year = {2021},
  month = {Dec},
  publisher = {American Physical Society},
  doi = {10.1103/RevModPhys.93.045003},
  url = {https://link.aps.org/doi/10.1103/RevModPhys.93.045003}
}

@misc{aceituno2024thermalization,
title = {{Thermalization and Localization: Novel Perspectives from Random Circuits and the Information Lattice (Ph.D. Thesis)}},
author = {Aceituno Ch\'avez, David},
year = {2024},
url={https://urn.kb.se/resolve?urn=urn:nbn:se:kth:diva-356733}
}

@article{zenodo_repo,
author  = {Bilinskaya, Yuliya and Martínez, Miguel F. and Ghosh, Soumi and Klein Kvorning, Tomas and Artiaco, Claudia and Bardarson, Jens H.},
  title  = {Witnessing Short- and Long-Range Nonstabilizerness via the Information Lattice},
  year         = {2025},
  journal    = {10.5281/zenodo.19852470},
  url          = {https://doi.org/10.5281/zenodo.19852470},
}
%################################

\end{document}